# Revision of the ionization energy of neutral carbon

W. L. Glab
Department of Physics and Astronomy, Texas Tech University, Lubbock, TX 79409, USA

K. Haris[1] and A. Kramida
National Institute of Standards and Technology, Gaithersburg, MD 20889, USA

## Abstract

This publication describes a re-analysis of previously published data on neutral carbon (C I) utilizing critically examined and improved values for the line energies of the absorption spectrum of molecular iodine to calibrate the transition energies of C I absorption lines originating from the $1s^22s^22p3s\ ^3P°_2$ level and terminating on highly excited Rydberg states of the $1s^22s^22pnp\ ^3D_3$ series, which converges on the first excited fine-structure level of the ionic ground configuration. Additional use of improved energy values for the $1s^22s^22p3s\ ^3P°_2$ level, the fine-structure interval of the ionic ground state, and sophisticated modern fitting techniques lead to a new value of the ionization energy of neutral carbon with 9 times the precision of previous work, 90820.348(9) cm$^{-1}$, as well as a table of predicted values for the energies of not yet observed states in the series.

**Keywords: neutral carbon, ionization energy, energy levels**

---

[1] Guest Researcher

## 1. Introduction

Accurate and precise values of photophysical quantities of atoms and molecules are important as input data in a number of fields, especially astrophysical calculations of properties, such as opacities, and chemical kinetics. These properties of the species involved include transition energies of both allowed and forbidden transitions, transition probabilities, dissociation energies of molecules, and ionization energies of the neutral and ionic species. A detailed review of the scope and importance of accurate photophysical data on carbon can be found in a recent paper by two of us [1].

Atomic carbon is a particularly important species in organic chemistry and astrophysics. In this brief article we describe a new determination of the ionization energy of neutral atomic carbon from the ground state, based on work by the group at NIST [1] and a related re-evaluation of data previously published by one of us [2].

A number of previous works have yielded ionization energy values with a variety of precisions [3, 4, 5]. As measurement technologies have become more refined, it is the norm for measured atomic properties to become more accurate over time. A common precise method of determining ionization energies is to measure the energies above the ground state for a large number of members of one or more Rydberg series of the species of interest. Preferably, the series used should be free of spin-orbit or other perturbations, eliminating the need for a multichannel quantum defect theory treatment of the data. The evolution of laser spectroscopic techniques, along with improved values of atlases of calibration lines [6, 7, 8] (in this case the absorption lines of molecular iodine) has made new measurements possible, and also enabled the re-evaluation of older data to a higher degree of precision.

## 2. Discussion of previously published results

In our 1998 work [2], we used nanosecond pulsed VUV+UV two photon resonant excitation to produce spectra of C I Rydberg atoms from $[1s^2 2s^2]2p^2\ {}^3P_0$ ground state atoms prepared by laser ablation, utilizing VUV resonance through selected fine structure levels of the $2p3s\ {}^3P°_J$ term followed by single-photon Rydberg state excitation using scanned wavelength tunable UV light from a frequency-doubled probe dye laser (which was energy calibrated at the fundamental wavelength of the probe laser by comparison to a simultaneously recorded spectrum of ground state molecular iodine). This scheme produced excitation selective in the total angular momentum $J$ of various Rydberg series converging to the ${}^2P°_{1/2}$ and ${}^2P°_{3/2}$ fine structure levels of the ground configuration $1s^2 2s^2 2p$ of the $C^+$ ion. Detection of the excited atoms was accomplished by collecting ions produced through either pulsed electric field ionization or autoionization. The ionization potential was determined through the use of the unperturbed $2pnp\,{}^3D_3$ series, which converges to the higher-energy ${}^2P°_{3/2}$ ionic fine-structure level. A simple fit of the observed transition energies to a Rydberg formula with the inclusion of an energy-independent quantum defect gave an interval between the intermediate state and the ${}^2P°_{3/2}$ series limit of 30490.54(3) cm$^{-1}$ with a quantum defect of 0.673(10). Added to the best available (at the time of the original work [2]) value of the intermediate $2p3s\ {}^3P°_2$ level energy [5], 60393.14(5) cm$^{-1}$, the energy required to ionize the ground state atom to an ion in the excited ${}^2P°_{3/2}$ state was found to be 90883.68(8) cm$^{-1}$. At our level of precision, subtracting the value of the ${}^2P°_{1/2} - {}^2P°_{3/2}$ energy splitting of the ionic ground configuration determined from far

IR laser magnetic resonance [9], 63.35 cm$^{-1}$, from our result for the $^2P°_{3/2}$ series limit did not increase the uncertainty and yielded a value of 90820.33(8) cm$^{-1}$ for the first ionization limit of C I. This was in reasonable agreement with the result of [5], but slightly more precise.

**3. Limitations of the accuracy of our previous result for the ionization energy**

It should be noted that the precision of our result for the ionization energy could be improved in several ways. First, the major part of our uncertainty came from the value of the energy of the intermediate resonance transition. Since the time of our work, improved values for the resonance line energies have become available, making a correction possible. Additionally, the group at NIST [1] derived accurate values for low 2*pnp* $^3D_3$ levels for *n* between 3 and 10. Including these levels in the fit would be expected to yield an improved series limit energy.

There are also several developments that allow us to improve the accuracy of our observed Rydberg state energies. Our original photon energy calibration was performed utilizing absorption lines energies of iodine from the atlas of Gerstenkorn and Luc [10] (henceforth referred to as GL). Since our earlier work, re-examination of the iodine spectrum [6, 7, 8] has shown that a weakly energy-dependent systematic correction factor (≈ –0.006 cm$^{-1}$ at the probe laser frequencies in our spectral region) must be applied to the energies of GL. Additional corrections to the assumed energies of the observed iodine lines used for our calibration have very recently been derived by one of us [H] by convolving our experimental probe laser bandwidth with these improved values of the iodine lines, including blended lines and hyperfine structure. While providing improved energies for many lines, this work also determined that some blended lines were unsuitable for calibration at our level of resolution; therefore, these lines were not used in the process of re-calibrating our Rydberg transition energies.

The scan rate that we used in our previous work was such that each iodine line had only 5 to 7 data points on it. This, combined with our noise level, made fitting of a Gaussian lineshape to each line to determine its center no more, or perhaps even less, accurate than "eyeball" weighting the data points of the lines to determine their centers. Errors introduced by using the visual technique were expected, and continue to be expected, to average out as random errors in the determination of the energies of a large number of Rydberg states.

A further difficulty that we discovered later was that our linear actuator driving the grating of the probe dye laser (a Thorlabs[1] piezoelectrically driven micrometer) would occasionally "glitch" in such a way that the energy scan rate in small regions of the spectrum (over a range of several wavenumbers) would deviate from its nominal value by about 20% over regions of several wavenumbers. Since we calibrated the spectrum in pieces by using the iodine lines in a number of regions of about 20 $cm^{-1}$ width each, fitting the observed iodine energies over each region to a polynomial function (generally second order), this led to relatively inaccurate values for several of the carbon Rydberg transition energies.

## 4. Recalibration of the Rydberg transition energies for the $^3D_3$ series and redetermination of the ionization energy

[1] Commercial products are identified in this article for adequate specification of the experimental procedure. This identification does not imply recommendation or endorsement by NIST.

Due especially to the scan rate problem noted above, it was decided to recalibrate the Rydberg transition energies on an individual basis, using 2 or 3 iodine lines with energies that bracketed the carbon transition energy and a linear fit of the iodine line energies to data point numbers to determine transition energies for each principal quantum number *n*. These fits yielded revised Rydberg state transition energies; additionally, each fit gave a scan rate that could be used to verify the choice of iodine lines and identify the regions where scan rate glitches occurred. Calibrations using the iodine lines that were determined by Haris to be unreliable were avoided. The estimated uncertainties in the Rydberg state energies derived from these fits intervals between the intermediate state and the Rydberg state were estimated to be approximately ±0.03 $cm^{-1}$ in the full double-frequency photon energy, based on the standard deviations of our measured wavenumbers from the fit of the Ritz formula described below.

Two of us [H and K] utilized all the available data (*n* = 3 to 10 from [1] and our present results for *n* = 40 to 69) in a four-term modified Ritz formula using the RITZPL [11] code to perform a fit yielding the series limit energy with its associated uncertainty. The residuals of this fit are shown in figure 1 below.

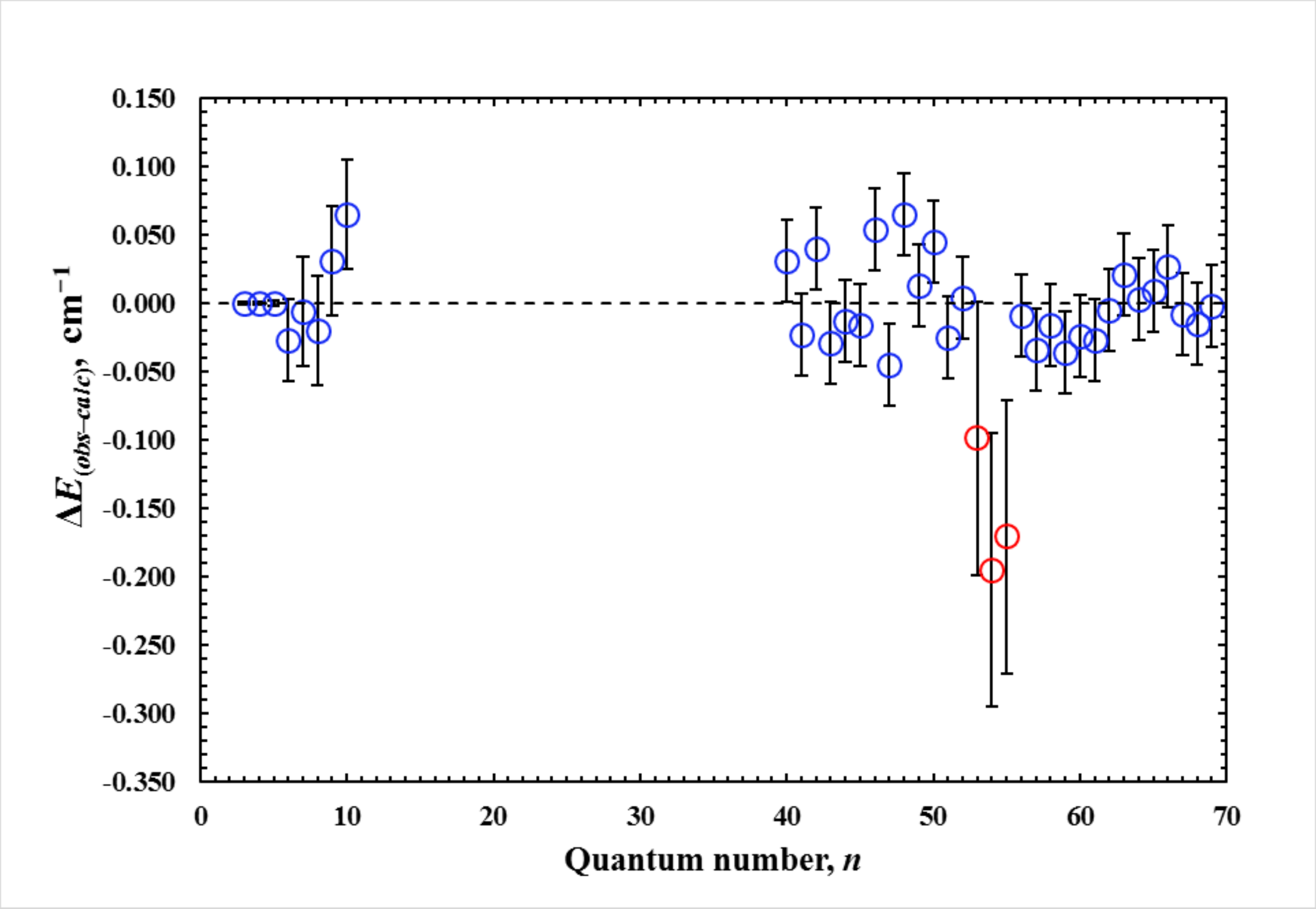

Figure 1. Residual differences $\Delta E_{(\text{obs-calc})}$ between the observed and fitted Rydberg energies. The error bars correspond to the measurement uncertainties.

The fit residuals are satisfactory with the exception of those for $n = 53$, 54, and 55. The energy region for these transitions suffered from both a scan rate glitch and a paucity of useful iodine calibration lines. After several unsuccessful attempts to calibrate these three lines, we concluded that they could not be calibrated to the accuracy of the other lines. They were therefore taken as outliers and removed from the data set for the fit for the series limit. Doing so led to a change in the resulting series limit barely within the final statistical error range of the ionization energy.

The fit was performed on a data set that consisted of the measured Rydberg state transition energies added to the intermediate state energy of 60393.1693(14) cm$^{-1}$ [1]. The fit with a 4-term modified Ritz formula, see Eq. (1b) in ref.[12], yields 90883.743(9) $cm^{-1}$ (with levels $n$ = 53, 54, and 55 excluded) for the interval between the atomic ground state and the ionic $^2P°_{3/2}$ excited state, where the uncertainty includes the following contributions: the standard deviation of the fit with the given values of energies, 0.005 $cm^{-1}$, the standard deviation (0.006 $cm^{-1}$) of 500 fits (of the same Ritz formula) with the level energies randomly varied around their nominal values with normal distributions of a width corresponding to their measurement uncertainties, and a systematic uncertainty in the measured $n \geq 6$ energies, estimated to be 0.003 $cm^{-1}$. It should be noted that the original RITZPL code of Sansonetti [11] was modified by one of us [K] for estimating the uncertainty by random trials. Subtracting from the limit value the ionic ground-state fine-structure interval from ref. [9], 63.39509(2) $cm^{-1}$ with more significant figures than in our earlier work gives the ionization energy of C I from the ground state to the ionic ground state as 90820.348(9) $cm^{-1}$. Table 1 below shows the results for observed and fitted values of the Rydberg energies above the ground state of the neutral atom, as well as the calculated quantum defects. Calculated values of unobserved energy levels are included, since they are accurate enough to be of use in future studies involving photoabsorption spectrum of atomic carbon. Uncertainties of the calculated energies $E_{calc}$ for $n < 6$ are equal to the standard deviations of 500 random trial fits for each level. For higher $n$, the above random-trial uncertainty estimate was combined in quadrature with the standard deviation of the fit for the limit value, 0.005 $cm^{-1}$, and an estimated systematic uncertainty of the observed $n \geq 6$ levels, 0.003 $cm^{-1}$.

**Table1.** Observed and Predicted Energy Levels of the $1s^22s^22pnp$ $^3D_3$ ($n$ = 3 to 69) series of neutral carbon

| $n$ | $E_{obs}$ [a] ($cm^{-1}$) | $E_{calc}$ [a] ($cm^{-1}$) | $\Delta E_{(obs-calc)}$ ($cm^{-1}$) | $\delta_{calc}$ [a] |
|---|---|---|---|---|
| 3 | 69744.0521(13) | 69744.0521(13) | 0.0000 | 0.721663250(22) |
| 4 | 80834.6249(13) | 80834.6249(13) | −0.0000 | 0.69551906(5) |
| 5 | 84985.0033(24) | 84985.0032(24) | 0.0001 | 0.68691915(14) |
| 6 | 87002.31(3) | 87002.337(17) | −0.027 | 0.6829267(15) |
| 7 | 88135.84(4) | 88135.846(20) | −0.006 | 0.6807295(25) |
| 8 | 88836.15(4) | 88836.170(18) | −0.020 | 0.679387(3) |
| 9 | 89299.13(4) | 89299.099(15) | 0.031 | 0.678505(3) |
| 10 | 89621.09(4) | 89621.025(13) | 0.065 | 0.677895(4) |
| 11 | | 89853.925(11) | | 0.677455(4) |
| 12 | | 90027.847(10) | | 0.677127(4) |
| 13 | | 90161.152(9) | | 0.676876(4) |
| 14 | | 90265.571(9) | | 0.676679(5) |
| 15 | | 90348.887(8) | | 0.676523(5) |
| 16 | | 90416.425(8) | | 0.676396(6) |
| 17 | | 90471.934(8) | | 0.676292(7) |
| 18 | | 90518.108(8) | | 0.676205(8) |
| 19 | | 90556.930(8) | | 0.676132(8) |
| 20 | | 90589.881(8) | | 0.676071(9) |
| 21 | | 90618.089(8) | | 0.676018(10) |
| 22 | | 90642.422(8) | | 0.675972(11) |
| 23 | | 90663.558(8) | | 0.675933(12) |
| 24 | | 90682.035(8) | | 0.675898(14) |
| 25 | | 90698.279(8) | | 0.675868(15) |
| 26 | | 90712.638(8) | | 0.675841(16) |
| 27 | | 90725.391(8) | | 0.675817(17) |
| 28 | | 90736.770(8) | | 0.675796(19) |
| 29 | | 90746.965(8) | | 0.675777(20) |
| 30 | | 90756.135(8) | | 0.675759(22) |
| 31 | | 90764.412(8) | | 0.675744(23) |
| 32 | | 90771.910(8) | | 0.675730(25) |
| 33 | | 90778.722(8) | | 0.67572(3) |
| 34 | | 90784.931(8) | | 0.67571(3) |
| 35 | | 90790.605(8) | | 0.67569(3) |
| 36 | | 90795.803(8) | | 0.67569(3) |
| 37 | | 90800.579(8) | | 0.67568(3) |
| 38 | | 90804.975(8) | | 0.67567(4) |
| 39 | | 90809.032(8) | | 0.67566(4) |
| 40 | 90812.82(3) | 90812.784(8) | 0.031 | 0.67565(4) |
| 41 | 90816.24(3) | 90816.260(8) | −0.023 | 0.67565(4) |
| 42 | 90819.53(3) | 90819.486(8) | 0.040 | 0.67564(4) |
| 43 | 90822.46(3) | 90822.487(8) | −0.029 | 0.67564(5) |
| 44 | 90825.27(3) | 90825.282(8) | −0.013 | 0.67563(5) |
| 45 | 90827.87(3) | 90827.890(8) | −0.016 | 0.67563(5) |
| 46 | 90830.38(3) | 90830.328(8) | 0.054 | 0.67562(5) |
| 47 | 90832.56(3) | 90832.609(8) | −0.045 | 0.67562(5) |
| 48 | 90834.81(3) | 90834.747(8) | 0.065 | 0.67561(6) |
| 49 | 90836.77(3) | 90836.754(8) | 0.013 | 0.67561(6) |
| 50 | 90838.69(3) | 90838.640(8) | 0.045 | 0.67561(6) |
| 51 | 90840.39(3) | 90840.415(8) | −0.025 | 0.67560(6) |
| 52 | 90842.09(3) | 90842.087(8) | 0.004 | 0.67560(7) |
| 53 | 90843.57(10) [b] | 90843.664(8) | −0.099 | 0.67560(7) |
| 54 | 90844.96(10) [b] | 90845.153(8) | −0.195 | 0.67559(7) |
| 55 | 90846.39(10) [b] | 90846.560(8) | −0.170 | 0.67559(8) |
| 56 | 90847.88(3) | 90847.893(8) | −0.009 | 0.67559(8) |
| 57 | 90849.12(3) | 90849.154(8) | −0.034 | 0.67559(8) |
| 58 | 90850.34(3) | 90850.351(8) | −0.016 | 0.67558(8) |
| 59 | 90851.45(3) | 90851.486(8) | −0.036 | 0.67558(9) |
| 60 | 90852.54(3) | 90852.564(8) | −0.024 | 0.67558(9) |
| 61 | 90853.56(3) | 90853.589(8) | −0.027 | 0.67558(9) |
| 62 | 90854.56(3) | 90854.565(8) | −0.005 | 0.67558(10) |
| 63 | 90855.52(3) | 90855.494(8) | 0.021 | 0.67557(10) |
| 64 | 90856.38(3) | 90856.379(8) | 0.003 | 0.67557(10) |
| 65 | 90857.23(3) | 90857.223(8) | 0.009 | 0.67557(11) |
| 66 | 90858.06(3) | 90858.029(8) | 0.027 | 0.67557(11) |
| 67 | 90858.79(3) | 90858.798(8) | −0.008 | 0.67557(11) |
| 68 | 90859.52(3) | 90859.534(8) | −0.015 | 0.67557(12) |
| 69 | 90860.24(3) | 90860.237(8) | −0.002 | 0.67556(12) |

[a] Observed energies ($E_{obs}$), calculated energies ($E_{calc}$), and quantum defects ($\delta$) for the $2s^2 2pnp\ ^3D_3$ ($n$ = 3 to 69) series, with their uncertainties (in parentheses, in units of the last decimal place of the value). All experimental energies, except for levels with $n \le 10$, are from the present work; $E_{obs}$ for $n \le 10$ are from Ref. [1]. In the fitting of the Ritz formula, the levels were weighted by inverse squares of their measurement uncertainties. See text for explanation of derivation of the $E_{calc}$ uncertainties.

[b] These levels were excluded from the fitting of the Ritz formula (see text).

## Conclusions

We have reanalyzed our previously published Rydberg energy measurements for C I through re-assessment of our observed 2*pnp* $^3D_3$ series Rydberg transition energies, utilizing critically analyzed and improved energy values for the molecular-iodine calibration lines to determine the Rydberg transition energies, and recently improved available data for associated energy intervals in the carbon atom. Our principal result is a new value for the ionization energy of the neutral carbon atom, 90820.348(9) $cm^{-1}$ or 11.2602880(11) eV in accordance with the 2014 CODATA recommended conversion factor for $cm^{-1}$ to eV [13], with a precision about 9 times better than our previously published result. Our analysis shows that, in order to produce the most accurate $I_2$-based calibrations possible with probe lasers of finite linewidth, a great deal of care must be taken to treat complications in the iodine calibration spectrum such as blended lines and unresolved hyperfine structure.

## Acknowledgements

Kramida and Haris were partially supported by the Astrophysics Research and Analysis program of the National Aeronautics and Space Administration of the USA. Haris was working at NIST under a Guest Researcher agreement 131227.

## References

1. Haris K and Kramida A 2017 *Astrophys. J. Suppl. Ser.* **233** 16
2. Glab W L, Glynn P T, and Robicheaux F 1998 *Phys. Rev. A* **58** 4014
3. Herzberg G 1958 *Proc. R. Soc. A* **248** 309
4. Kaufman V and Ward J F 1966 *J. Opt. Soc. Am.* **56** 1591

5. Johansson L 1966 *Ark. Fys.* **31** 201
6. Kato H, Kasahara S, Misono M, and Maasaki B 2000 *Doppler-Free High Resolution Spectral Atlas of Iodine Molecule 15,000 To 19,000cm$^{-1}$* (Tokyo: Japan Society for the Promotion of Science) and accessible from **http://hdl.handle.net/1811/19890**
7. Knöckel H, Bodermann B, and Tiemann E 2004 *Eur. Phys. J. D* **28** 199
8. Salami H and Ross A J 2005 *J. Mol. Spectrosc.* **233** 157
9. Cooksy A L, Blake G A, and Saykally R J 1986 *Astrophys. J.* **305** L89
10. Gerstenkorn S and Luc P 1979 *Atlas Du Spectre D'absorption de la Molecule D'iode 14800–20000 cm$^{-1}$* (Paris: Laboratoire Aimé Cotton CNRS II)
11. Sansonetti C J 2005 *Computer programs RITZPL* private communication
12. Kramida A 2013 *Fusion Sci. Technol.* **63** 313
13. Mohr P J, Newell D B, and Taylor B N 2016 *J. Phys. Chem. Ref. Data* **45** 043102